\documentclass[pre,twocolumn,showpacs]{revtex4}
\usepackage{graphicx}
\usepackage{amsbsy,amssymb,amsmath,bm,ulem,float}

\usepackage{color}

\graphicspath{{../fig/}}

\begin{document}

\title{Dendritic flux avalanches in rectangular superconducting films -- numerical simulations}

\author{J. I. Vestg{\aa}rden}
\affiliation{Department of Physics, University of Oslo, P. O. Box
1048 Blindern, 0316 Oslo, Norway}
\author{Y. M. Galperin}
\affiliation{Department of Physics, University of Oslo, P. O. Box
1048 Blindern, 0316 Oslo, Norway}
\author{T. H. Johansen}
\affiliation{Department of Physics, University of Oslo, P. O. Box
1048 Blindern, 0316 Oslo, Norway}
\affiliation{Institute for Superconducting and Electronic Materials,
University of Wollongong, Northfields Avenue, Wollongong,
NSW 2522, Australia}
\affiliation{Centre for Advanced Study at The Norwegian Academy
of Science and Letters, Drammensveien 78, 0271 Oslo, Norway}

\begin{abstract}
Dendritic flux avalanches is a frequently encountered instability in
the vortex matter of type II superconducting films at low
temperatures.  Previously, linear stability analysis has shown that
such avalanches should be nucleated where the flux penetration is
deepest.  To check this prediction we do numerical simulations on a
superconducting rectangle. We find that at low substrate temperature
the first avalanches appear exactly in the middle of the long
edges, in agreement with the predictions. At higher substrate
temperature, where there are no clear predictions from the theory, we
find that the location of the first avalanche is decided by
fluctuations due to the randomly distributed disorder.
\end{abstract}

% 74.25.Ha  Magnetic properties
% 74.62.En  Effects of disorder
% 74.78.-w  Superconducting films and low-dimensional structures
% 74.25.Fy      Transport properties
% 74.25.Qt Vortex lattices, flux pinning, flux creep
% 74.25.fc     Thermal conduction in SC
% 68.60.Dv Thermal stability; thermal effects
% 74.78.-w  Superconducting films and low-dimensional structures
\maketitle

% n0100 Ha = 0.052 Jc0
% n0125 Ha = 0.065 Jc0
% n0150 Ha = 0.078 Jc0
% n0184 Ha = 0.14 Jc0

%keywords: Dendritic flux avalanches; thermomagnetic instability; flux jump; flux creep simulation;

\section{Introduction}
Dendritic flux avalanches have been observed in many kinds of type~II
superconducting films at low temperatures~\cite{altshuler04}.
The origin of the avalanches is a thermomagnetic instability mechanism
between the Joule heating created by vortex motion and the reduction
of the critical current density as temperature
increases~\cite{mints81,rakhmanov04,denisov05}.

When transverse applied field is gradually increased, a critical state is formed 
from the edges, with almost constant current density, and non-zero magnetic flux density.
The flux penetration is gradual and smooth until the conditions for 
onset of instability is fulfilled. Then, an avalanche is nucleated and 
large amounts of magnetic flux rushes in
from the edges and forms a complex tree-like structure. The onset
conditions for such events are determined by a competition between the
Joule heating, heat diffusion and heat removal to the
substrate~\cite{denisov05}. From this theory, it is expected that 
instabilities are most likely to happen where the electric field is
highest and the flux penetration is deepest.  Since rectangles in low
applied fields experience a flux penetration which is slightly deeper
in the middle of the long edges~\cite{brandt95}, 
we hence expect the first avalanches to be nucleated there. 
At higher fields, the flux front straightens out, and it is less 
clear where the most favorable nucleation location will be.

In this work, we will run numerical simulations of a
thermomagnetically unstable superconducting rectangle in applied
transverse field. 
The location of the avalanches will be discussed in 
context of prediction from linear stability analysis. The
simulation method is described in Ref.~\cite{vestgarden11}.

\section{Model}
Consider a type II superconducting film subjected to an applied 
magnetic field transverse to the film plane, 
ramped at a constant rate $\dot H_a$. The shape of the 
sample is a rectangle with dimensions $2a\times 2b$.

A thermomagnetic instability is a consequence of the 
nonlinear material characteristics of type II superconductors, 
which conventionally is approximated by a power law
\begin{equation}
  \label{EJ}
  \mathbf E = \frac{\rho_0}{d}\left(\frac{J}{J_c}\right)^{n-1}\mathbf J
  ,
\end{equation}
where $\mathbf E$ is electric field, $\mathbf J$ is sheet current, $J=|\mathbf J|$,
$\rho_0$ is a resistivity constant,  $d$ is sample thickness, 
$J_c$ is critical sheet current, and $n$
is the creep exponent. The temperature dependencies are taken as 
\begin{equation}
  J_c=J_{c0}(1-T/T_c),~~~ n=n_1/T,
\end{equation}
where $T_c$ is the critical temperature. The electrodynamics must be supplemented 
by the heat diffusion equation 
\begin{equation}
  \label{dotT}
  c\dot T = \kappa \nabla^2T-\frac{h}{d}(T-T_0)+\frac{1}{d}JE
  ,
\end{equation}
where $c$ is specific heat, $\kappa$ is thermal conductivity, $h$
is the coefficient for heat removal to the substrate, 
and $T_0$ is the substrate temperature. 

Eq.~\eqref{dotT} must be solved together with Maxwell's equations and
the material law, Eq.~\eqref{EJ}.  The description of the simulation
method is in Ref.~\cite{vestgarden11} and here only the key
ingredients are outlined.  The main challenge is to invert the
Biot-Savart law in an efficient way. This is done by including also
the vacuum outside the sample in the simulation formalism. At the cost
of including the extra space, one can use a real space/Fourier space
hybrid method with the very attractive performance scaling of $O(N\log
N)$, where $N$ is the number of grid points.

Quenched disorder is important in order to give realistic
nucleation conditions, and it is introduced as a $5\%$ reduction of
$J_{c0}$ in randomly selected $5\%$ of the grid points.

\begin{figure}[t]
  \centering
  \includegraphics[width=\columnwidth]{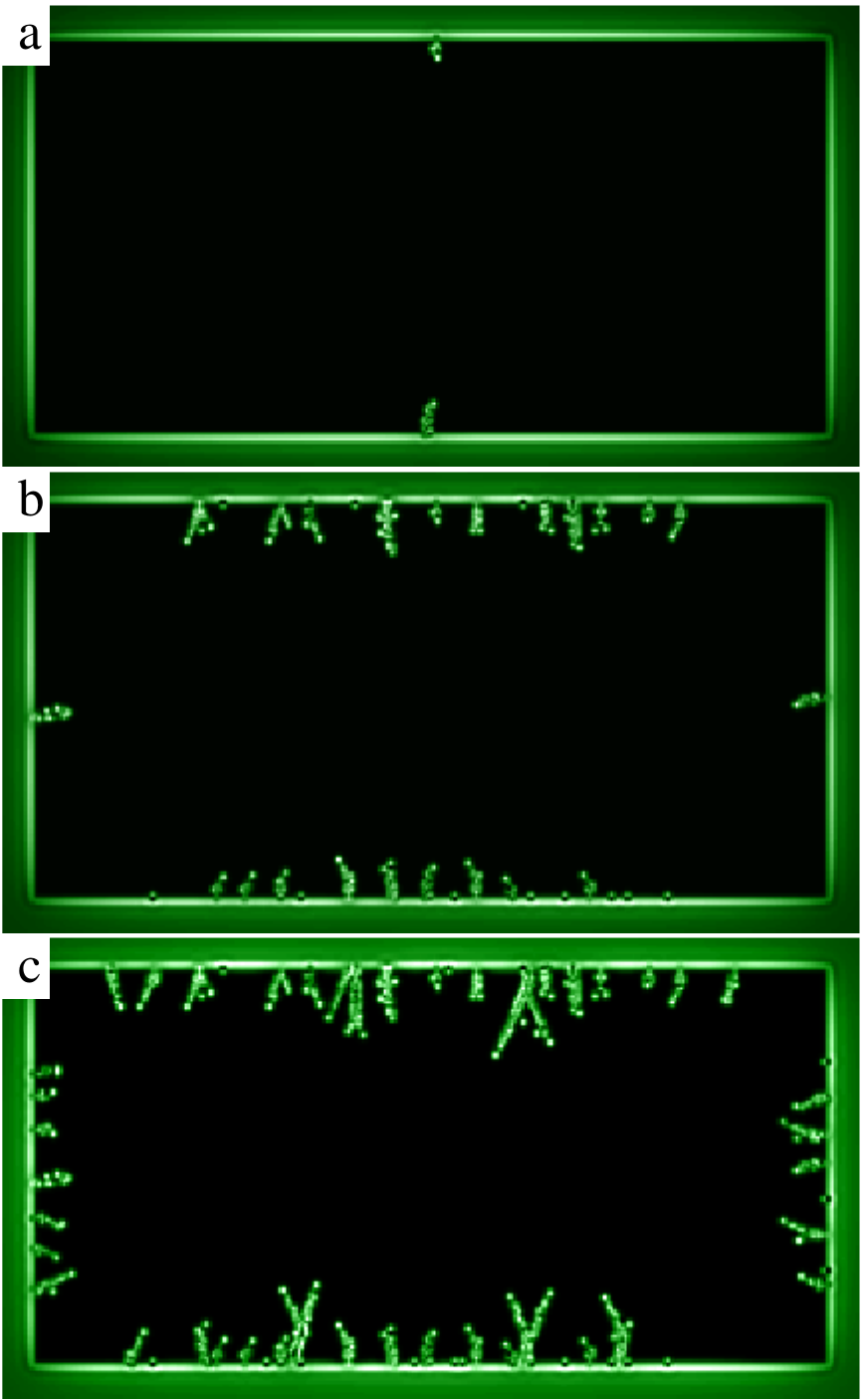} 
  \caption{
    Distributions of magnetic flux density, $B_z$, 
    at $H_a/J_{c0}=0.052$(a), 0.064(b), and 0.078(c). 
    Substrate temperature is $T_0=0.17T_c$. 
    The magnetic field is highest at the edges, seen
    as a white rim, while the black central region is still flux-free, i.e., $B_z=0$.    
    The avalanches are all small, most of them are fingers, while in (c)
    some avalanches also have two or three branches.
    \label{fig:B17}
  }
\end{figure}

\begin{figure*}[t]
  \centering
  \includegraphics[width=18cm]{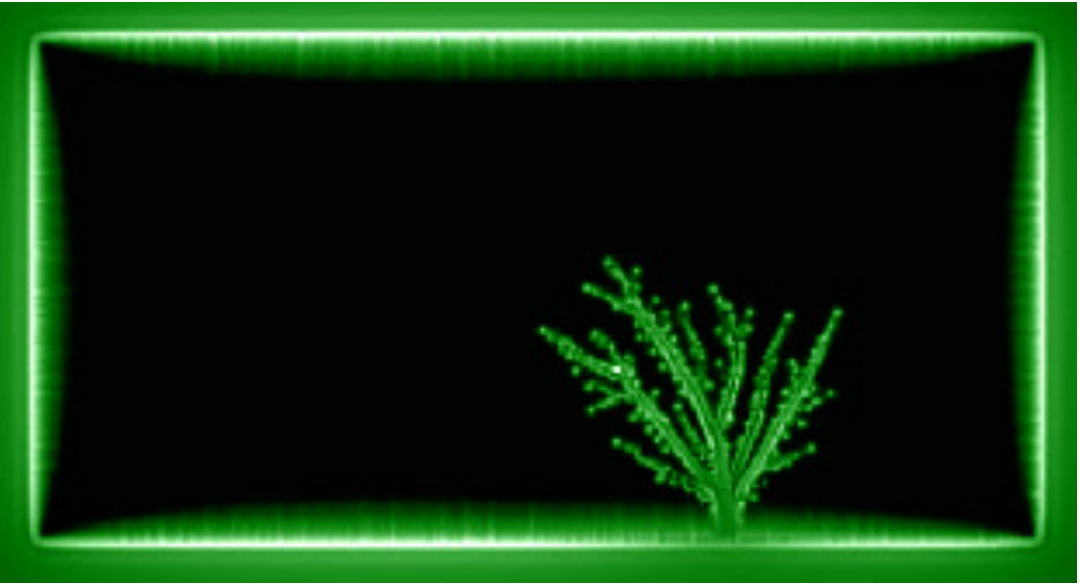} \\
  \caption{
    \label{fig:B22}
    Distribution of magnetic flux density, $B_z$, at $H_a/J_{c0}=0.14$ with
    substrate temperature $T_0=0.22T_c$.
    Only one dendritic avalanche has been nucleated. Note that the location is not in 
    the middle of the side and the avalanche is 
    larger and have more branches than those 
    at lower $T_0$, in Fig.~\ref{fig:B17}.  
  }
\end{figure*}

The parameters correspond to typical 
values for MgB$_2$ films \cite{schneider01}:
$T_c = 39$~K,
$j_{c0} = 1.2\times 10^{11}$~A/m$^2$,
$\rho_n = 7~\mu\Omega$cm,
$\kappa = 170$~W/Km$\times (T/T_c)^3$,
$c = 35$~kJ/Km$^3\times (T/T_c)^3$,
$h = 200$~kW/Km$^2\times (T/T_c)^3$,
%$\dot B_a = 10$~T/s, 
and
$n_1=15$. 
Here $\rho_n$ is normal resistivity, and we have 
$J_{c0}=dj_{c0}$, and $\rho_0=\rho_n$.
The sample dimensions are
$2a = 8~$mm,
$2b = 4~$mm,
and 
$d = 0.4~\mu$m,
while the total simulated area is $12\times 6$~mm$^2$,
where the extra space outside the sample is used 
to implement the boundary conditions. 
The total area is discretized on a $768 \times 384$
equidistant grid.

\section{Analytical prediction}
The linear stability analysis in Ref.~\cite{denisov06} determines 
the condition for instability onset as 
\begin{equation}
  l^*=\frac{\pi}{2}\sqrt{\frac{d\kappa}{|J_c'|E}}\left(1-\sqrt{\frac{h}{n|J_c'|E}}\right)^{-1}
\end{equation}
where $l^*$ is the threshold flux penetration depth,
$|J_c'|=J_{c0}T_0/T_c$, and $E$ is the background electric field.  For
a given $E$, the sample is unstable when the flux front exceeds $l^*$,
which means that avalanches should appear first where the flux penetration is deepest. 
In a rectangle, this is in the middle of the
two long sides and consequently it is expected that the first avalanches
appear there.

\section{Simulation results}
The simulations start from initially zero-field cooled state, and the
applied field is increased at constant rate, $\mu_0\dot H_a=10~$T/s.
The high ramp rate was chosen by performance reasons, since a high
ramp rate closes the gap between the velocity of the avalanches and
the normal flux penetration. Yet, there is clear separation of time
scales, since full penetration is reached in approximately
$J_{c0}/\dot H_a=6$~ms, while the duration of the avalanches is less
than $0.1~\mu$s.

Fig.~\ref{fig:B17} shows the distributions of magnetic 
flux density transverse to the film plane, $B_z$, at $H_a/J_{c0}=0.053$(a),
0.064(b), and 0.078(c).  The substrate temperature is $T_0=0.17T_c$,
and at this low temperature the threshold field for avalanche
activity is low, $H_\text{th}=0.052J_{c0}$.  Image (a) is just above
the threshold, where two small fingers have appeared symmetrically
in the middle of the two long edges, just as expected from the
theory. In image (b) many more avalanches have come at the long sides,
and for the first time there are avalanches appearing at the two
short sides, also these in the middle. In image
(c) there are even more avalanches and they now cover most of the
boundary, except close to the corners.  All avalanches are small, and
will consequently be seen as small, jumps towards zero in
magnetization curves. With increasing field, it seems like a
trend of increasing avalanche size.

Fig.~\ref{fig:B22} shows $B_z$ for a simulation run at a higher
substrate temperature, $T_0=0.22T_c$.  The figure contains just one
avalanche, which appeared at 
$H_\text{th}=0.12J_{c0}$. Both the increased $H_\text{th}$ and the
much larger size of the avalanche is as expected for higher
$T_0$~\cite{denisov06}.  Also expected is the complex
branching pattern~\cite{vestgarden11}. The location is not entirely symmetric. The reason
is that at deeper flux penetrations the flux front straightens out, as
seen in regular Bean-state of the upper part of the image. Hence, the
location of the avalanche is instead determined by the fluctuations
in electric field due to the quenched disorder. The avalanche
location is typical for what is seen experimentally, e.g., by
magneto-optical images in NbN rectangles~\cite{yurchenko07}.

\section{Conclusions}
We have simulated dendritic flux avalanches in superconducting 
films in the shape of a rectangle. The results confirm the prediction
from linear stability analysis that avalanches will first appear in
the middle of the long sides, at least for low $T_0$. At higher $T_0$,
the avalanche location was not symmetric.  In general, the results
regarding avalanche size, threshold field, and time between avalanches
follow the typical dependency on $T_0$, i.e., that increasing $T_0$
gives larger $H_\text{th}$, larger avalanches, more branches, and
longer time between the avalanches.  Hence, the simulations of this
work also confirms the correctness of the simulation method of 
Ref.~\cite{vestgarden11}.

\bibliographystyle{elsart-num}
%\bibliography{../bibtex/superconductor}
%\bibliography{superconductor}

\end{document}